\documentstyle[12pt]{article}
\renewcommand{\baselinestretch}{1.5}
\renewcommand{\arraystretch}{1.5}
\begin{document}
\thispagestyle{empty}
\pagestyle{empty}
\renewcommand{\thefootnote}{\fnsymbol{footnote}}
\newcommand{\preprint}[1]{\begin{flushright}
\setlength{\baselineskip}{3ex}#1\end{flushright}}
\renewcommand{\title}[1]{\vspace{1ex} \begin{center}\Large #1\end{center}\par}
\renewcommand{\author}[1]{\vspace{2ex}{\it \begin{center}
\setlength{\baselineskip}{3ex}#1\par\end{center}}}
\renewcommand{\thanks}[1]{\footnote{#1}}
\renewcommand{\abstract}[1]{\vspace{2ex}\normalsize\begin{center}
\centerline{\bf Abstract}\par\vspace{2ex}\parbox{6in}{#1
\setlength{\baselineskip}{2.5ex}\par}
\end{center}}
\newcommand{\starttext}{\newpage\normalsize
\pagestyle{plain}
\setlength{\baselineskip}{4ex}\par
\setcounter{footnote}{0}
\renewcommand{\thefootnote}{\arabic{footnote}}
}
\newcommand{\segment}[2]{\put#1{\circle*{2}}}
\newcommand{\jpsi}{J \! / \! \psi}
\newcommand{\fig}[1]{figure~\ref{#1}}
\newcommand{\ds}{\displaystyle}
\newcommand{\eqr}[1]{(\ref{#1})}
\newcommand{\be}{\begin{equation}}
\newcommand{\ee}{\end{equation}}
\def\lte{\mathrel{\displaystyle\mathop{\kern 0pt <}_{\raise .3ex
\hbox{$\sim$}}}}
\def\gte{\mathrel{\displaystyle\mathop{\kern 0pt >}_{\raise .3ex
\hbox{$\sim$}}}}
\newcommand{\sechead}[1]{\medskip{\bf #1}\par\bigskip}
\newcommand{\ba}[1]{\begin{array}{#1}\ds }
\newcommand{\cra}{\\ \ds}
\newcommand{\ea}{\end{array}}
\newcommand{\bra}[1]{\left\langle #1 \right|}
\newcommand{\ket}[1]{\left| #1 \right\rangle}
\newcommand{\braket}[2]{\left\langle #1 \left|#2\right\rangle\right.}
\newcommand{\braketr}[2]{\left.\left\langle #1 right|#2\right\rangle}
\newcommand{\g}[1]{\gamma_{#1}}
\newcommand{\half}{{1\over 2}}
\newcommand{\del}{\partial}
\newcommand{\grad}{\vec\del}
\newcommand{\real}{{\rm Re\,}}
\newcommand{\imag}{{\rm Im\,}}
\newcommand{\gapprox}{\raisebox{-.2ex}{$\stackrel{\textstyle>}
{\raisebox{-.6ex}[0ex][0ex]{$\sim$}}$}}
\newcommand{\lapprox}{\raisebox{-.2ex}{$\stackrel{\textstyle<}
{\raisebox{-.6ex}[0ex][0ex]{$\sim$}}$}}
\newcommand{\cl}[1]{\begin{center} #1\end{center}}
\newcommand{\dzero}{D\O}
\newcommand{\etal}{{\it et al.}}
\newcommand{\prl}[3]{Phys. Rev. Letters {\bf #1} (#2) #3}
\newcommand{\prd}[3]{Phys. Rev. {\bf D#1} (#2) #3}
\newcommand{\npb}[3]{Nucl. Phys. {\bf B#1} (#2) #3}
\newcommand{\plb}[3]{Phys. Lett. {\bf #1B} (#2) #3}
\newcommand{\ie}{{\it i.e.}}
\newcommand{\etc}{{\it etc.\/}}
\renewcommand{\baselinestretch}{1.5}
\renewcommand{\arraystretch}{1.5}
\newcommand{\boxit}[1]{\ba{|c|}\hline #1 \\ \hline\ea}
\newcommand{\mini}[1]{\begin{minipage}[t]{20em}{#1}\vspace{.5em}
\end{minipage}}
\preprint{MSUHEP-93-01}
\preprint{January 6, 1993}
\title{
Double Pomeron Opportunities at $\sqrt{s} = 1.8 \, {\rm TeV}$
\thanks{Research supported in part by Texas National
Research Laboratory Commission grant RGFY9240 to the CTEQ Collaboration.}
}
\author{
Jon Pumplin \\
Department of Physics and Astronomy \\
Michigan State University \\
East Lansing MI 48824 \\
{\footnotesize Bitnet: PUMPLIN@MSUPA}
}
%\date{}
\abstract{
I describe possible ways to discover hard double pomeron exchange
(HDPE) with the existing detectors at the Fermilab Tevatron, by
using the small-angle ``luminosity'' counters as a veto.  Estimates
of the cross sections and backgrounds are made.  In addition to the
intrinsic importance of HDPE, its observation would be useful for
calibrating the detectors, and for estimating the
``survival probability'' of rapidity gaps.
}

\starttext

\section{Introduction}
\label{sec:introduction}
Typical ${\rm \bar p p}$ interactions at
$\sqrt{s} = 1.8 \, {\rm TeV}$ produce a large number of particles,
which are distributed rather uniformly in pseudo-rapidity
$\eta = - \ln \tan {\theta \over 2 \,}$, with $dN/d\eta \sim 6$ for
$\vert \eta \vert < 4$.  There can nevertheless exist final states
with one or more {\it ``rapidity gaps''\/}, defined as intervals of
length $\Delta \eta > 2$ to $3$ containing zero particles.  Long
rapidity gaps are by definition governed by the {\it pomeron}, which
is believed related to the $s$-channel unitarity phenomenon of shadow
scattering.  A QCD-based understanding of the pomeron remains elusive,
although a qualitative description as a two-gluon system with vacuum
quantum numbers is promising\cite{lownuss,pumplin,landshoff,FS}.

The `grand-daddy' of rapidity gap processes is elastic scattering,
which makes up $\sim 20 \%$ of $\sigma_{\rm tot}$.  Inelastic single
diffraction, defined by a gap with a leading ${\rm p}$ or
${\rm \bar p}$ at one end, is also responsible for a sizeable fraction
of $\sigma_{\rm tot}$, and has been seen to exhibit hard-scattering
effects\cite{UA8}.  Double pomeron exchange (DPE), defined by
{\it two} rapidity gaps, has been observed using special detectors
for small angle quasi-elastic protons at
$\sqrt{s} = 0.063 \, {\rm TeV}$\cite{ISR} and
$\sqrt{s} = 0.63 \, {\rm TeV}$\cite{Kernan}.

A complete study of rapidity gap physics demands detectors that cover
a long range in $\eta$, to establish the absence of particles in one
or more gap regions while detecting particles outside the gaps.  Such
detectors have been proposed for
SSC ($\sqrt{s} = 40 \, {\rm TeV}$)\cite{FAD} and the FNAL
Tevatron ($\sqrt{s} = 1.8 \, {\rm TeV}$)\cite{MAX}.  The point of
the present paper, however, is to consider some
{\it hard double-pomeron exchange (HDPE)} processes that can be
studied at the Tevatron in the working detectors CDF\cite{CDF} and
\dzero\cite{D0}, which cover roughly $-4 < \eta < 4 \,$.  The processes
are defined by rapidity gaps of $-4 < \eta < -2$ and $2 < \eta < 4$,
with an experimentally clean high-$Q^2$ object in the central region.
That object should be producible from two pomerons, which are assumed
to have vacuum quantum numbers.

Promising candidates for the central object are
(1) Pairs of $b \bar b$ bound states:  $\Upsilon(1S) \Upsilon(1S)$
or $\Upsilon(2S) \Upsilon(2S)$;
(2) Single $b \bar b$ bound states:  $\chi_{b0}(1P)$, $\chi_{b0}(2P)$,
$\chi_{b2}(1P)$, or $\chi_{b2}(2P)$; or
(3) Two jets separated by $\Delta \eta < 1$.
The odd-C $b \bar b$ states can be detected by
$\Upsilon \rightarrow e^+ e^-$ or $\mu^+ \mu^-$.
The even-C states can be detected by
$\chi \rightarrow \gamma \Upsilon$ followed by
$\Upsilon \rightarrow e^+ e^-$ or $\mu^+ \mu^-$\cite{PDG}.

If any of these $\Upsilon \Upsilon$ or $\chi_b$ final states are
observed, the DPE nature of their production can be demonstrated by
showing that the cross section does not decrease drastically when the
required rapidity gaps are extended into the central rapidity region.
There is plenty of room for this in the case of the $b \bar b$ states.
For the 2-jet final states, the requirement $\Delta \eta < 1$ also
leaves some room for extending the rapidity gaps into the central region
to make this test.  One can also check that single $\Upsilon$ states are
strongly suppressed relative to $\chi_b$, since they cannot be made from
two pomerons according to charge conjugation\cite{odderon}.

We will show that the above processes offer a reasonable opportunity to
observe the hard scattering of two pomerons (HDPE).  The $\Upsilon$ and
$\chi_b$ states are especially attractive because their $b \bar b$ wave
functions are relatively well understood, which will facilitate attempts
to compute the production.  Because of their low multiplicity and
accurately known masses, these states will also be valuable
experimentally for calibrating the energy resolution and noise level
of the detector.

\section{Trigger and backgrounds}
\label{sec:trigger}

At current Tevatron luminosities, interactions occur at a rate of a few
$\times \, 10^5$ per second, while events can be recorded at a rate of
a few per second.  Background events must therefore be rejected by
$\sim \! 10^{-5} \,$.  The trigger decision is made in a series of stages,
with the initial rejection of $10^{-2}$ to $10^{-3}$ based on rather
incomplete information.  I propose to cope with this trigger challenge
as follows.

A minor part of the \dzero detector consists of scintillation counters
that cover approximately $-4 < \eta < -2$ and $2 < \eta < 4$.  Similar
counters cover $3.24 < \vert \eta \vert < 5.90$ in CDF.  Normal
triggers require a coincidence between hits in these ``luminosity''(\dzero)
or ``beam-beam''(CDF) counters to make a first estimate of the
interaction vertex by timing, and to discriminate against beam+gas
interactions.  {\it The crucial experimental proposal I make is to use
the luminosity counters instead as a veto.}  A trigger defined by an
absence of hits in these counters, in coincidence with some indication
of hard scattering in the central region $\vert \eta \vert < 2$, will
eliminate a large fraction of ``ordinary'' events to look for HDPE.

To study the trigger, I use the QCD Monte Carlo program
{\footnotesize HERWIG5.5}\cite{herwig} to simulate events.  Events that
contain no charged particles in the veto regions $-4 < \eta < -2$ and
$2 < \eta < 4$ are analyzed in a simple model of a calorimeter detector,
consisting of cells $0.20 \times 0.20$ in $\eta$ $\times$ azimuthal
angle $\phi$.   Requiring a transverse energy $E_T > 2.5 \, {\rm GeV}$
or an electromagnetic transverse energy
$E_T^{\rm EM} > 1.5 \, {\rm GeV}$ in at least one of the cells in the
central region, and no electromagnetic or hadronic energy above
$0.1 \, \rm GeV$ in the veto regions, I find a cross section of
$\sim 1 \, {\rm \mu b}$ according to the ``minimum bias'' mode
of {\footnotesize HERWIG}.

A second opinion on the minimum bias physics can be obtained from
the $2 \rightarrow 2$ QCD hard scattering mode of
{\footnotesize HERWIG}, with the ``background event'' turned off and
the minimum transverse momentum in hard scattering set to a small
value ($2.2 \, {\rm GeV}$) chosen to produce the entire inelastic
cross section.  This ``mini-jet'' model predicts a similar rate for
the electromagnetic $E_T$ trigger, and a somewhat larger rate
$\sim 2 \, {\rm \mu b}$ for the hadronic trigger.  Leaving the
``background event'' on would give a lower rate.

It will be better to trigger on coincidences between two cells above
an $E_T$ threshold in the central region.   This will further reduce
the trigger rate from minimum bias physics, and at the same time
suppress non-physics backgrounds from detector noise, beam halo,
and beam-gas collisions.  In this way it will be possible to look
for HDPE processes, which are of course not simulated by
{\footnotesize HERWIG}, down to the smallest cross sections visible
at the Tevatron luminosity.

\section{Estimates of signals}
\label{sec:estimates}

I estimate diffractive $\Upsilon \Upsilon$ production by a
pole-dominance model, used long ago to calculate DPE $\pi^+ \pi^-$
production\cite{henyey}.  The amplitude from Fig.~1 is
\begin{equation}
{\cal M} =
T(p_1 \, q \rightarrow p_3 \, p_4) \;
(q^2 - M_{\Upsilon}^2)^{-1} \;
[1 - A(q^2 - M_{\Upsilon}^2)]^{-1} \;
T(p_2 \, - \! q \rightarrow p_6 \, p_5)
\label{eq:eq11}
\end{equation}
An exchanged graph is obtained by $p_4 \leftrightarrow p_5$.
The two-body $\Upsilon {\rm p}$ elastic amplitudes are
\begin{equation}
T(p_1 \, q \rightarrow p_3 \, p_4) \, = \,
i \, s_{34} \, \sigma_{\Upsilon p} \, e^{\beta \, t_{13} / 2}
\label{eq:eq12}
\end{equation}
where $s_{34} = (p_3 + p_4)^2$ and $t_{13} = (p_1 - p_3)^2$.
Reasonable guesses for the $\Upsilon p$ total cross section and
forward elastic slope are $\sigma_{\Upsilon p} = 2 \, {\rm mb}$
and $\beta = 6 \, {\rm GeV}^{-2}$.
Eq.(\ref{eq:eq11}) includes an off-shell suppression factor
controlled by the parameter $A$, which is hard to guess but
expected to be ${\cal O} (1 \, {\rm GeV}^{-2})$.  For
$A = 1 \, {\rm GeV}^{-2}$, this model gives $4 \, {\rm nb}$ for
$\Upsilon \Upsilon$ production.  Allowing for the branching
fractions $\Upsilon \rightarrow  e^+  e^-$ or $\mu^+  \mu^-$ for
both of two $\Upsilon(9460)$s, and making the rapidity and $E_T$
cuts above reduces the estimate to $4 \, {\rm pb}$.  {\it At the
anticipated Tevatron integrated luminosity of $50 \, pb^{-1}$,
this will lead to 200 events} and be clearly visible.

The pole-dominance model predicts that DPE production of
$\jpsi \, \jpsi$ will also be observable.   Assuming
$\sigma_{\jpsi \, {\rm p}} = 3 \, {\rm mb}$ and again using
$A = 1 \, {\rm GeV}^{-2}$ leads to
$20 \, {\rm nb}$ for $\jpsi \, \jpsi$ production.  When both
$\jpsi$ decay to $e^+  e^-$ or $\mu^+  \mu^-$, at least one of
the four leptons has  $E_T > 3 \, {\rm GeV}$ more than half
of the time.  This relatively large $E_T$ arises because the
individual $\jpsi$ transverse momenta are comparable
to $M_{\jpsi}$, even though their sum is small.  The DPE
$\jpsi \jpsi$ leptonic final states will therefore also generally
pass our proposed trigger.

We next attempt to estimate the diffractive production of single
$\chi_{b}$ states.  First note that these states can be formed by
gluon+gluon fusion with a coupling strength that is measured by
their hadronic width:
\begin{equation}
\hat \sigma_{g g \rightarrow \chi} =
(\pi^2 \, \Gamma_{\chi \rightarrow g g } / 16 M_\chi) \,
\delta(\hat s - M_\chi^2) \quad ,
\label{eq:eq13}
\end{equation}
which includes a factor $1/256$ from spin and color averaging.
The state $\chi_{b0}(2P)$  has $M_{\chi} = 10.23 \, {\rm GeV}$
and $\Gamma_{\rm hadronic} \approx 400 \, {\rm KeV}$\cite{PDG}.
According to a simple parton-model calculation, it is formed by
$g+g$ fusion, with a production cross section integrated over
$\vert \eta \vert < 1.5$ of $\sim \! 20 \, {\rm nb}$.  Including
the poorly known branching ratios to $\gamma \Upsilon$ followed
by $\Upsilon \rightarrow e^+ e^-$ or $\mu^+ \mu^-$ reduces the
observable cross section for this {\it non-pomeron} process to
$\sim 30 \, {\rm pb}$.

One can imagine a second gluon exchange that modifies the
$g+g$ fusion process and makes the overall exchange between beam
and target a color singlet.  This color singlet exchange
does not necessarily produce a large average number of particles
per unit of rapidity.  It occasionally produce zero particles
in the veto regions $-4 < \eta < -2$ and $2 < \eta < 4$.   If the
price for the two gaps is less than a factor $\sim 1/300$, the
DPE production rate of the $\chi_b$ state will be observable.
For a discussion of this idea, see reference\cite{BBH}.

A similar estimate can be made for
$g + g \rightarrow {\rm jet} + {\rm jet}$.
{}From {\footnotesize HERWIG}, the cross section for two jets with
$E_{T} > 10 \, {\rm GeV}$, $\vert \eta \vert < 1.3$, and
$\vert \eta_1 - \eta_2 \vert < 1.0$ is
$\, \sim 3 \times 10^7 \, {\rm pb} \,$.  If a second gluon exchange
can produce rapidity gaps as suggested above, the corresponding HDPE
process will be observable even if the gap requirement suppresses
this huge cross section by $10^{-8}$.

\section{Conclusion}
\label{sec:conclusion}

I have shown that a trigger based on two rapidity gaps can be used to
look for hard double pomeron interactions in the existing detectors at
the Tevatron.  Trigger rates and non-diffractive backgrounds are small
enough to look for these processes down to $\sim 0.1 \, {\rm pb}$.

Crude estimates suggest that several HDPE processes will be observable.
I have emphasized processes with one or two $b \bar b$ states, because
they have experimentally very clean $e^+ e^-$ and $\gamma e^+ e^-$
decay modes with sufficient transverse energy to satisfy the trigger.
The $b \bar b$ states are also attractive theoretically because one can
use their known wave functions in attempting to calculate the
production.  Final states involving two $\jpsi$ instead of two
$\Upsilon$, with nothing else visible in the detector, are also worth
looking for.

Final states with two jets nearby in rapidity, with gaps on either side,
will allow the most sensitive search for HDPE, since two-jet production
must have a relatively large cross section compared to the other
processes we consider.  Requiring $\vert \eta_1 - \eta_2 \vert < 1.0$
for the jet axes, with jets defined by cones
$\sqrt{\Delta \eta^2 + \Delta \phi^2} < 0.7$, leaves an average
of $2.8$ units in $\eta$ on either side of the ${\rm jj}$ system
to define the rapidity gaps.  Assuming one observes HDPE candidates,
in which no particles (or no calorimeter cells above the noise level)
appear in the gap regions, it will be important to study the
distribution in the number of particles in the gap regions.  One must
see if there is a peak at 0 particles which signals HDPE; or if the
0-particle events appear to be simply fluctuations of normal hard
scattering.  This background due to fluctuation has a cross section
$\sim 0.3 \, {\rm nb}$ for jets defined by $E_T > 10 \, {\rm GeV}$,
according to a {\footnotesize HERWIG} simulation.

Observing HDPE processes would also establish the survival of rapidity
gaps, which offer important possibilities for Higgs and $\rm W W$
scattering physics at the SSC\cite{higgs}.

\section*{Acknowledgements}
I thank H. Weerts for discussions on \dzero and J. Huston
for discussions on CDF.

%\newpage

%\newpage

\begin{center}
FIGURE CAPTION
\end{center}

\begin{enumerate}

\item Pole model for DPE production of $\Upsilon \Upsilon$ or
$\jpsi \jpsi$.  The blobs represent elastic
scattering, which is dominated by pomeron exchange.

\end{enumerate}
\end{document}